# The Composability of Intermediate Values in Composable Inductive Programming

Edward McDaid FBCS
Chief Technology Officer
Zoea Ltd

Sarah McDaid PhD
Head of Digital
Zoea Ltd

**Abstract**

It is believed that mechanisms including intermediate values enable composable inductive programming (CIP) to be used to produce software of any size. We present the results of a study that investigated the relationships between program size, the number of intermediate values and the number of test cases used to specify programs using CIP. In the study 96,000 programs of various sizes were randomly generated, decomposed into fragments and transformed into test cases. The test cases were then used to regenerate new versions of the original programs using Zoea. The results show linear relationships between the number of intermediate values and regenerated program size, and between the number of test cases and regenerated program size within the size range studied. In addition, as program size increases there is increasing scope for trading off the number of test cases against the number of intermediate values and vice versa.

**1. Introduction**

Research into various forms of inductive programming has been carried out over several decades [1,2,3,4,5,6] however until recently there has not been a serious attempt to produce a general purpose programming language that incorporates this technology. The main reason for this is the fact that inductive programming systems are still only capable of directly producing relatively small programs - the equivalent of a few lines of code in a conventional programming language [7]. It is natural to wonder whether inductive programming on its own will ever be able to produce programs of arbitrary size.

Over the same period the evolution of programming languages has continued apace. Many thousands of programming languages have been created and many programming paradigms have been identified [8]. Yet all mainstream programming languages remain fundamentally imperative in nature [9]. Even putatively declarative languages require the developer to specify the operations involved although they can be less specific about the order in which these need to happen. Contemporary programming languages often attempt to support multiple paradigms and have a tendency to grow in size over time [9]. As a result all programming languages are complex and learning to code is difficult.

**2. Composable Inductive Programming**

Composable inductive programming (CIP) is an approach to software development that uses inductive programming to generate code from a specification that resembles test cases [10]. In addition CIP allows generated code to be combined (or composed) in various ways to form larger programs. Zoea [10] and Zoea Visual [11] are two examples of CIP languages. Zoea is a text based language that resembles YAML [12] but consists mostly of data in a format that is a superset of JSON [13]. Zoea Visual is a visual programming language built on top of Zoea that introduces the concept of dependencies to support the creation of data flow test case diagrams. Both Zoea and Zoea Visual are simple languages and enable the definition of programs with much lower complexity than conventional languages [14].

Zoea programs consist of a set of one or more test cases. Each test case normally has an input and an output value. Zoea supports two forms of composition. The first is expressed as any number of intermediate values between the input and output values of a test case. In Zoea intermediate values are called derived values. A derived value between an input and an output is treated internally as two sequential back-to-back test cases - i.e. input-derive and derive-output. Intermediate values allow developers to articulate data transformations deep within the code path for a test case. Each derived or output value is also called a step.

Zoea also supports composition at the program level through the 'use' relationship. When we define a program in Zoea we can suggest that it incorporates (or uses) one or more existing Zoea programs. This causes the Zoea compiler to temporarily treat the specified existing programs as additional instructions in the instruction set for its internal virtual machine. Zoea will therefore try to incorporate the code for the existing programs into the solution for the new program - if it can. This is similar to the concept of import in many conventional programming languages but without any specific indication of where in the solution the imported code should actually be used. Zoea programs that use other programs can form a hierarchy of any size.

Zoea Visual also supports intermediate values and the use relationship but introduces an additional form of composition called subsidiary test cases. A subsidiary test case is an embedded test case diagram that represents a piece of code in a similar way to intermediate values. Unlike intermediate values the subsidiary test case can be highly complex and internally can comprise any number of steps and dependencies. It can also have any number of test cases and can use completely different values from the test case in which it is embedded. This allows complex logic to be expressed with a linear rather than exponential number of test cases. Subsidiary test cases can also form a hierarchy of any size.

It is believed that composition enables CIP to be used to create programs of any size [10]. This is because composition can be used in conventional programming languages to combine small fragments of code such as methods and classes into very large codebases. Absolute limits such as maximum class size in Java [15] do exist in conventional languages but are rarely encountered by developers in practice.

Given that CIP languages and the CIP paradigm are very different from their conventional programming predecessors it would be useful to have more empirical evidence that composition in CIP works beyond simple examples. Also, there is currently little experience in the use of CIP as a software development approach. It would be useful to understand how different numbers of intermediate values affect the numbers of test cases required to produce a program and so on.

## 3. Study Approach

The high level objective of the study was to obtain and analyse statistics relating to program size, number of intermediate values and number of test cases for a set of programs produced using CIP. The high level strategy to achieve this was to produce many programs of different sizes with different numbers of intermediate values and different numbers of test cases and to look for any patterns that might be present in the data.

CIP is a new approach and at the time of writing CIP languages are not yet generally available. As a result relatively little Zoea code exists beyond regression tests and examples - most of which are small and focussed on demonstrating specific language features. Also, there are currently very few software developers with any practical experience in CIP.

Given the study would require many programs and that a significant proportion of these programs would need to be non-trivial in terms of size an approach based on the random generation of programs was adopted. Random generation also helps to reduce any risk of selection or confirmation bias that might arise in programs that were otherwise specified and coded manually.

Figure 1 provides an overview of the process used in the study. Program generation was carried out by a separate application which was specifically built for the purpose. In order to minimise any risk of cross-talk with the Zoea compiler the only existing software element from the Zoea codebase that was reused was the Zoea Internal Language (ZIL) interpreter. This was used to run randomly generated programs and to obtain output results from test cases. The ZIL code that Zoea produces internally is very similar to conventional programming languages in many respects.

Program generation simply used a list of ZIL instruction names and the number of arguments that each of these requires. It also used a random data set that included a large number of random data values of each data type. Random values were used as constant instruction arguments in generated code and also as the input values for generating test cases.

Assembly of each random program was initiated with a budget which is an integer representing the maximum Halstead length [16] of the required program (total number of operators + total number of operands). Instructions and arguments were selected randomly and the process also invoked itself recursively to produce nested code. At each stage the budget was decremented as appropriate and the process was complete when the budget was exhausted. The simple way in which this was implemented meant that generated programs were often smaller but never larger than the specified budget.

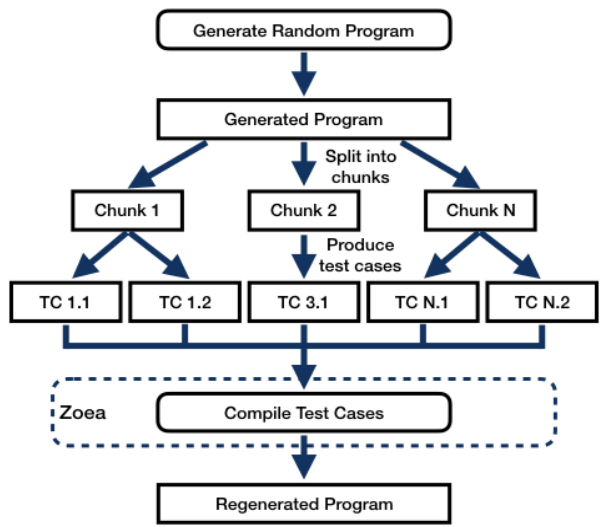

**Figure 1 – Study Approach Overview**

Each random program that was produced was evaluated to determine whether it worked. A program was deemed to work if:

- It could be executed by the ZIL interpreter;
- No exceptions were thrown;
- It produced results for a set of random inputs;
- The results were not always the same;
- At least one of the results was different from the corresponding input;
- The results were returned within 200 milliseconds.

The last condition filtered programs that looped indefinitely. Not surprisingly most programs generated in this way did not work at all but the process was still able to produce hundreds of working randomly generated programs every hour.

Each working program was then split into a random number of between 1 and N chunks where N is the minimum of 32 or the number of instructions the program contains. Each such chunk corresponds to the code that would be produced by an intermediate value or output in a Zoea test case.

For each code chunk a random number of between 1 and 8 test cases were produced by feeding random values into the chunk and noting the output if any. This process is similar to fuzz testing [17]. The same criteria that were applied to determine whether a program worked were also applied to the code chunks during test case creation. If the production of test cases was not successful for any chunk then the process was abandoned for that generated program and restarted with a different random program.

The resulting set of test cases for each randomly generated program were compiled using the Zoea compiler to produce a recreated version of the original randomly generated program. It is notable that the regenerated version is frequently smaller as the randomly generated programs often contain redundant instructions. For this reason only the size of the regenerated version is used and the length of the original random program is disregarded.

The randomly generated programs, the code chunks into which they were decomposed and the regenerated programs included both sequential and parallel data flow paths in virtually any combination. In the regenerated programs the overall data flow topology had no significant impact on the total program size. This is because the total program size is dominated by the sum of the sizes of the individual chunks regardless of topology.

If we forget about how we came by the test cases then this process is logically equivalent to simply using Zoea to produce a program of size S given a number of derived values D and a total number of test cases N. Aside from any of the benefits arising from randomisation the randomly generated programs simply served as an expedient way to specify a large number of non-trivial Zoea programs.

## 4. Study Results

The data output of the study related to 96,000 programs of different sizes and different total number of test cases for each of between 1 and 32 steps. This corresponds to 3000 different programs of different sizes and different numbers of test cases that were produced for each number of intermediate values. For each regenerated program the program size in bytes, the total number of test cases used to regenerate the program and the number of steps was recorded.

The largest program generated was 3263 bytes in size while the median size was 826 bytes and the standard deviation was 538.61 bytes.

Due to the random and unconventional way in which programs were constructed the distribution of the regenerated program sizes was not uniform. It is worth noting that the generated programs do not contain any formatting or white space that would often be present in human originated source code.

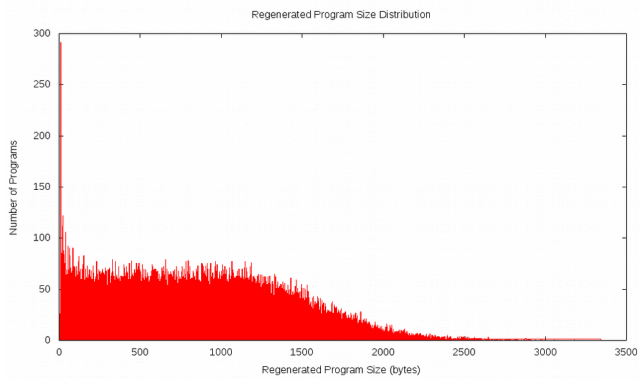

**Figure 2 – Regenerated Program Sizes**

Figure 2 shows a plot of the number of regenerated programs against regenerated program size in bytes. It can be seen that many very small programs were produced and few larger programs. Random generation of larger programs takes much more time than does generation of smaller programs. Also, it is more likely that a larger random program will be defective for some reason than a smaller one.

While the sizes of the randomly generated programs were specified to some extent the sizes of the regenerated programs were not. As we have already noted there were significant variations between the sizes of the generated programs due to factors such as redundant instructions. In addition there are often many different ways to express the same logic in code. The Zoea compiler will generally identify the most concise solution possible.

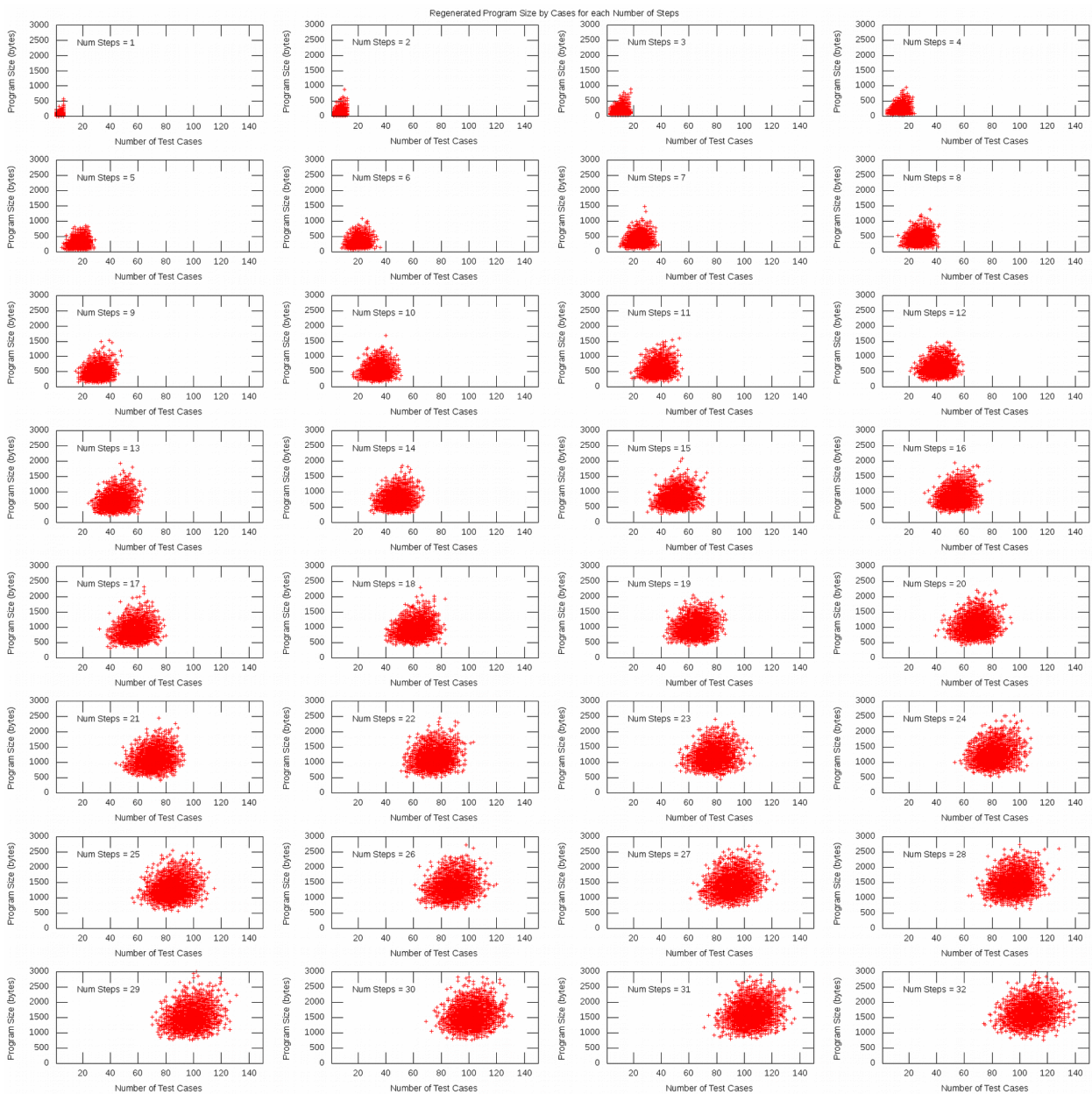

**Figure 3 – Regenerated Program Size by Test Cases for each Number of Steps**

The distribution of program size in the results corresponds to the regenerated programs as these are what the test cases and intermediate values actually specify in Zoea. For these reasons the distribution of program size is neither flat nor an inverse distribution as would normally be expected. It would have been possible to force the distribution to be more even in terms of program size however this would have required considerably more time than was available. It would also have involved discarding many working programs of small and intermediate size.

Program size is measured in bytes. There is no accurate way to translate this measurement into commonly used metrics such as lines of code as any such mapping must take account of the programming language used, data element sizes as well as developer formatting preferences. In approximate terms the largest programs generated in this study correspond to conventional programs that are well over 100 lines long.

Figure 3 shows the results as a series of plots of regenerated program size against total number of test cases for each number of steps. Each of these individual plots includes 3000 separate data points. Generally it can be seen that program size increases as the number of steps and the number of test cases increases. It can also be seen that the dispersion of the individual points increases as the number of steps and the number of test cases increase.

Figure 4 summarises the results in a single plot by showing median regenerated program size as a colour coded scatter plot against the number of steps and total number of test cases. The overall trend of increasing program size with increased number of steps and test cases is also apparent. It can be seen for example that with 10 steps we have results for programs specified by between approximately 20 and 50 test cases. On average this corresponds to between 2 and 5 test cases per step specified. The actual numbers of test cases for each step that go to make up each program is of course more complicated. In this study we are only considering the total number of test cases for each regenerated program.

Figure 5 provides a three dimensional plot of the study results. This plot uses a weighted average of any points with the same number of steps and test cases. The apparent peaks on the right are the result of outliers and relative sparsity of data rather than being actual features in the data.

Figures 6 and 7 show separate plots for regenerated program size against total number of test cases and regenerated program size against number of intermediate values respectively. The X-axis on both of these plots has been randomly jittered in order to allow better visibility of individual samples and also to provide a better indication of sample density where samples would otherwise be overlaid. An approximately linear relationship can be seen in both cases. Both plots also exhibit heteroscedasticity or non-constant variance which is a common characteristic of datasets in which values can have large ranges [18]. The presence of heteroscedasticity compromises the ability to carry out regression analysis which assumes that variance is constant. This can be corrected through a number of strategies including mapping the data or weighted least squares regression.

In this study the presence of increasing variance is not surprising given the combinatorial nature of program structure and composition. Large programs are made up of many smaller elements which can vary considerably in size. As programs grow in terms of their Halstead length the potential variability in their overall size in bytes becomes ever larger.

The data for figures 6 and 7 were mapped using the approach described in [18]. The Y-axis was transformed using each of log, square root and reciprocal. Of these square root achieved the greatest consistency in terms of variance while all of the transforms removed linearity. The same process was repeated with the X-axis values with the objective of restoring linearity. Again the square root transform yielded the best results in both cases. Figures 8 and 9 show the transformed results for program size against the number of test cases and program size against the number of steps respectively. These are clearly linear and have consistent variance. The Pearson correlation coefficients are 0.917 for program size by cases and 0.920 for program size by steps – both of which represent very strong positive associations.

## 5. Discussion and Future Work

The key finding of this study is confirmation of the intuition that increasing numbers of intermediate values can be used with CIP to create correspondingly larger resulting programs – at least within the program size range for which results were obtained. The study also confirmed a similar relationship between the number of test cases and program size.

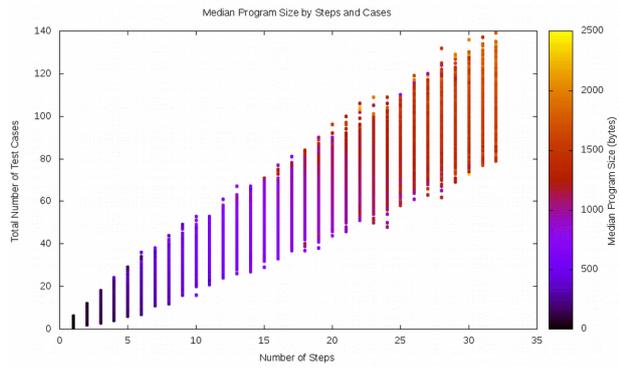

**Figure 4 – Median Program Size by Steps and Cases**

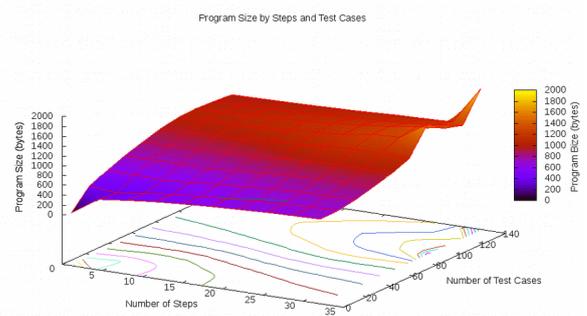

**Figure 5 – Program Size by Steps and Cases 3-D**

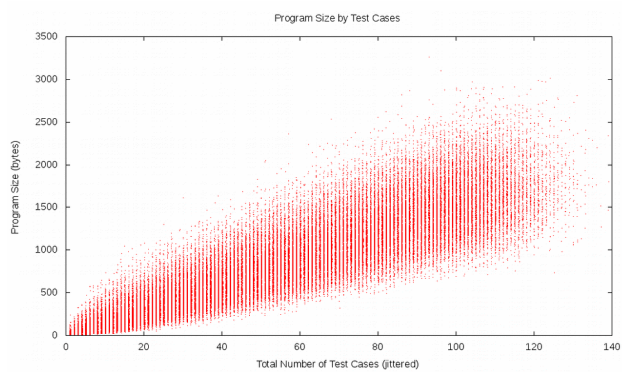

**Figure 6 – Program Size by Test Cases**

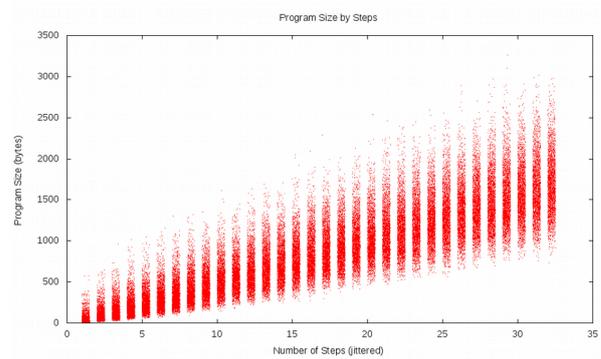

**Figure 7 – Program Size by Steps**

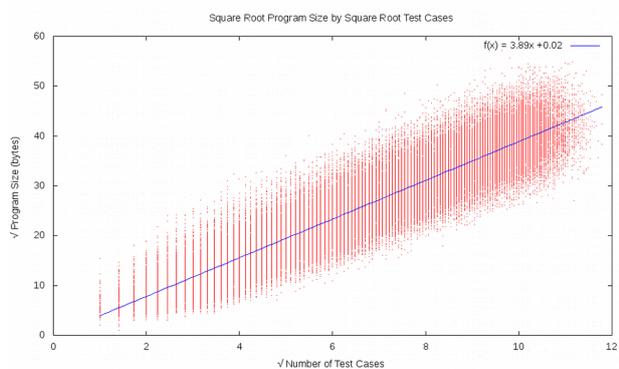

**Figure 8 – Modified Program Size by Test Cases**

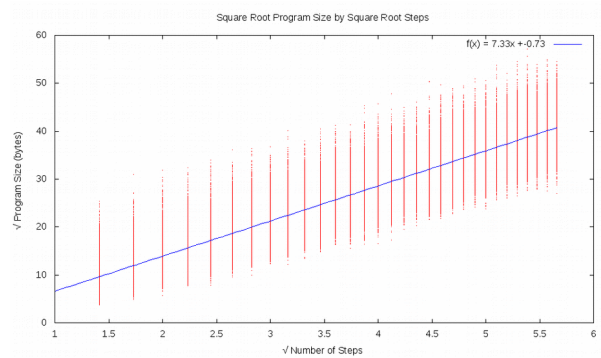

**Figure 9 – Modified Program Size by Steps**

If the number of intermediate values is fixed then increasing the number of test cases supports the creation of larger programs. Similarly if the number of test cases is fixed then increasing the number of intermediate values can also support the creation of larger programs.

The results are not a total surprise however before the study was conducted the nature of the relationship between test case numbers, intermediate values and program size was not known with certainty.

The numbers of test cases or steps involved in the production of the larger programs in this study might seem high but in practise these numbers should be lower. No attempts were made in the study to minimise the number of test cases nor the number of steps required to specify a given program. Instead the results reflect the sizes of programs that are possible to generate with sets of largely random test cases. Also, the Zoea programs that were generated in the study do not utilise other composition mechanisms such as uses or subsidiary test cases that would normally be used in programs of any significant size. In this regard the Zoea programs that were produced are somewhat atypical but this does not detract from the results as the study was designed to investigate a single form of composition.

The findings are also applicable as a lower bound to subsidiary test cases in Zoea Visual. Where subsidiary test cases are simply used with identical data values to the parent diagram they are equivalent in their expressive power to intermediate values. However, subsidiary test cases when used with different values are a much more powerful construct that can significantly reduce the overall number of test cases required. A separate study would be necessary to investigate this further but it would be expected that subsidiary test cases support the generation of larger programs with fewer cases and that this relationship would also be at least linear.

The current study could be extended in a number of ways. In particular it would be interesting to investigate still larger regenerated programs together with greater numbers of test cases and intermediate values. The main impediment to this would be the time and resources required for the random generation and compilation of many very large programs. One possible strategy to overcome these obstacles would be to create a very large set of small program fragments from which any number of large programs could be assembled. This approach would effectively enable us to extrapolate the results to virtually any program size with ease. Care would need to be taken to ensure sufficient diversity in generated programs however such a strategy would require each code fragment to be compiled by Zoea only once.

The current study has considered only the total number of test cases required to regenerate programs across any number of steps. This ignores a level of detail where each step has an associated number of test cases and each regenerated program has a distribution of test case numbers across all steps. Perhaps there is more insight to be had here.

Similarly we have not considered the size, makeup or information content of the test cases themselves. Experience with Zoea Visual suggests that where test cases include compound data structures such as arrays a smaller number of test cases is often required in order to specify a program of a given complexity. It is apparent that the expressive power of a set of test cases as a program specification is closely related to the number of data elements that it contains, as well as its Shannon [19] or Kolmogorov complexity [20]. The broader role of information theory in the formulation of test cases is an interesting topic and there would seem to be considerable scope for further work in this area [21].

We do not expect the overall data flow topology of the regenerated program to influence the results. This could also be confirmed experimentally.

The current study has not considered the 'use' form of composition in Zoea in any way. Investigating this would present some challenges. In order to do so we would have to be able to define programs $P_{parent}$ and $P_{child}$ in such a way that $P_{parent}$ can use $P_{child}$ in its compilation when suggested - either always or at least with a non-zero probability. If we can produce a solution for $P_{parent}$ without the need for use composition then any code fragment from the solution code for $P_{parent}$ could represent the code for a possible $P_{child}$. However, at some point sufficiently large or complex programs built using CIP will require use composition in the first place in order to produce any solution. Alternatively, if we take a bottom-up approach we can generate random code for $P_{child}$ and then use it to attempt to produce a random $P_{parent}$ that uses $P_{child}$ – repeating this process until we are successful. Using this latter approach it would be possible to iteratively produce ever larger random programs through use composition with any number of different used programs at each level. This should be able to demonstrate that very large programs can be constructed with CIP through use composition. At each stage we could also produce a set of test cases for the generated program. These could then be used to attempt to regenerate an equivalent program without use composition. This would help us to understand the size limits of programs that could be built without use composition.

## 6. Conclusions

We have presented the results of a study that investigated the relationships between the number of intermediate values in a CIP specification, the total number of test cases and size of the resulting program. In the study 96,000 random programs were generated, decomposed and transformed into sets of test cases. The test cases were then used to regenerate new versions of the original programs using the Zoea compiler. The study has shown that linear relationships exist between the number of intermediate values and program size, and between the number of test cases and program size. For a given program size if more intermediate values are used then fewer test cases are required and vice versa. We have also discussed the results and provided some suggestions for related future work.

**Acknowledgements**

This work was supported entirely by Zoea Ltd (https://www.zoea.co.uk). Zoea is a trademark of Zoea Ltd. All other trademarks are the property of their respective owners. Copyright © Zoea Ltd. 2021. All rights reserved.